\documentclass[conference]{IEEEtran}
\IEEEoverridecommandlockouts
\usepackage{cite}
\usepackage{amsmath,amssymb,amsfonts}
\usepackage{algorithmic}
\usepackage{graphicx}
\usepackage{textcomp}
\usepackage{xcolor}
\def\BibTeX{{\rm B\kern-.05em{\sc i\kern-.025em b}\kern-.08em
    T\kern-.1667em\lower.7ex\hbox{E}\kern-.125emX}}
\begin{document}

\title{Machine Learning-Enabled Cyber Attack Prediction and Mitigation for EV Charging Stations\\
}

\author{\IEEEauthorblockN{Mansi Girdhar and Junho Hong}
\IEEEauthorblockA{\textit{Electrical and Computer Engineering} \\
\textit{University of Michigan-Dearborn}\\
Dearborn, USA \\
gmansi@umich.edu and jhwr@umich.edu}
\and
\IEEEauthorblockN{Yongsik Yoo and Tai-jin Song}
\IEEEauthorblockA{\textit{Department of Urban Engineering} \\
\textit{Chungbuk National University}\\
Cheongju, South Korea \\
yys@chungbuk.ac.kr and tj@chungbuk.ac.kr}


}

\maketitle

\begin{abstract} 
Safe and reliable electric vehicle charging stations (EVCSs) have become imperative in an intelligent transportation infrastructure. Over the years, there has been a rapid increase in the deployment of EVCSs to address the upsurging charging demands. However, advances in information and communication technologies (ICT) have rendered this cyber-physical system (CPS) vulnerable to suffering cyber threats, thereby destabilizing the charging ecosystem and even the entire electric grid infrastructure. This paper develops an advanced cybersecurity framework, where STRIDE threat modeling is used to identify potential vulnerabilities in an EVCS. Further, the weighted attack defense tree approach is employed to create multiple attack scenarios, followed by developing Hidden Markov Model (HMM) and Partially Observable Monte-Carlo Planning (POMCP) algorithms for modeling the security attacks. Also, potential mitigation strategies are suggested for the identified threats.

\end{abstract}


\section{INTRODUCTION}
The global auto industry is undergoing a seismic transition from internal combustion engine dominance to electric vehicles (EVs) as a source of salvation for climate change. The penetration of EVCSs in the electric grids is increased due to the continuously growing fleet of EVs. Since internet-connected EVCS is a principal infrastructure in an intelligent power system, the cybersecurity perspective of EVCS is critical to the operation of power systems. 
Cyber attack patterns on critical infrastructures, especially energy delivery systems, are evolving and diversifying, indebted to the internet of things (IoT) paradigm, which has infused numerous vulnerabilities. Due to the complex cyber-physical interactions and interdependencies at the nexus of the EVs, EVCSs, and power grid, the exchanged data might be subject to disparate vulnerabilities. The perpetrators might use these existing vulnerabilities of the EVCS equipment to compromise the system, thus jeopardizing its security. 
The severe implications incurred by the adversarial attacks triggered a wide interest of white-hat hackers and researchers to address the cybersecurity aspects of the EVCSs. 
Although extensive research has been done on developing efficient and resilient cybersecurity frameworks for power grid applications, for e.g., development of Critical Infrastructure Protection (CIP) 002-009 standard by the North American Electric Reliability Corporation (NERC)~\cite{4753673}, and IEC 62351 standard by the International Electrotechnical Commission (IEC) Technical Committee (TC) 57 Working Group (WG) 15~\cite{9361308}, designing of anomaly detection systems (ADSs) in~\cite{6786500},~\cite{9306781} and~\cite{7553505}, and several others, but they do not reinforce the cybersecurity of EVCSs.
 
Also, EVCSs have not yet encountered distinguished and high-profile cyber incidents, and investigators have been interposing with the charging operations to determine the existence of unknown vulnerabilities.
For e.g., Kaspersky Lab researchers exposed the software vulnerabilities of EVCSs, which make them accessible to unauthorized hackers\cite{9272723}. Therefore, these known vulnerabilities, and more eminently a possibility of zero-day vulnerabilities, underline the cyber risks of EVCSs. 

Inspite of the fact that there have been some diligent efforts done to address the critical cybersecurity gaps in an EVCS, for e.g., the proposal of security standards for EV charging architectures by the European Network of Cybersecurity~\cite{7807218} and cybersecurity recommendations by the US DOE, NHTSA and, DOHS, these proposed recommendations are yet to be standardized and operational. Due to the unacceptable consensus for the EV charging protocols, the non-standard cyber-physical interfaces make EVs and EVCS susceptible to attacks that can damage the EVs and EVCS equipment. 

This paper emphasizes the cybersecurity concerns of an EVCS along with the detection and mitigation measures to counter the coordinated cyber attacks. The principal contributions of the paper are: (1) a cybersecurity framework which encompasses STRIDE-based threat modeling for an EVCS to identify potential cyber vulnerabilities, weighted attack defense tree to outline multiple cyber attack scenarios, HMM to predict the most likely path in a multi-stage attack and POMCP algorithm to decoy the attacker towards the predicted path of attack.



The remainder of the paper is divided as follows:
Section II discusses the cyber-physical model of an EVCS. Sections III outlines the holistic cybersecurity framework, including the applications of STRIDE threat modeling, weighted attack defense tree, HMM, and POMCP algorithms to analyze innumerable cybersecurity threats faced by an EVCS and generate various attack scenarios in the EVs and EVCS. Section IV evaluates the performance of the proposed model by a case study of denial of service (DoS) attack along with some potential mitigation strategies in section V. Finally, Section VI concludes the paper along with the limitations and recommendations for future work.

\section{HIGH-LEVEL OVERVIEW OF AN EVCS}
\begin{figure}[t]
\centering
\includegraphics[width= .5\textwidth, height = 2.6 in]{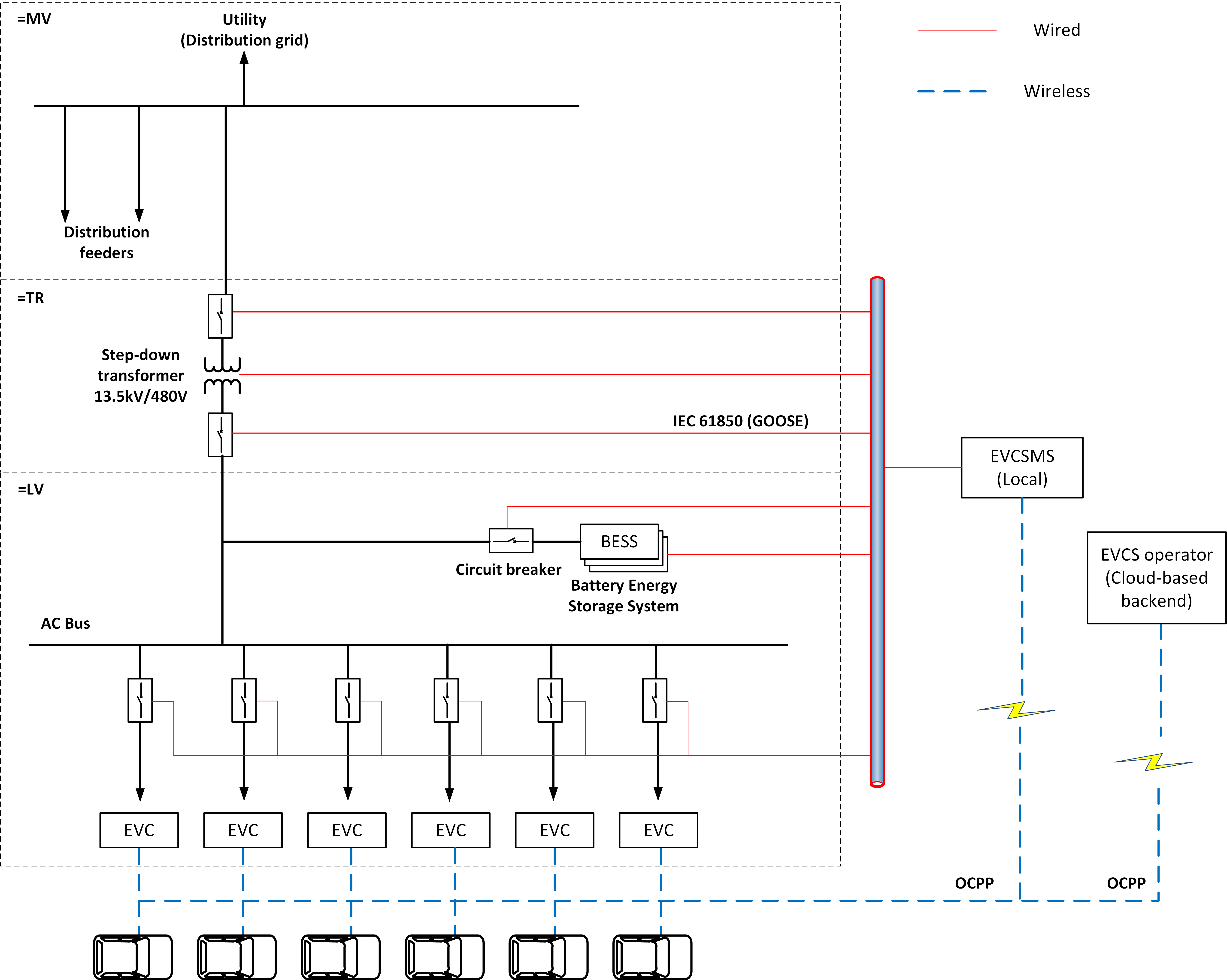}
\caption{High-level overview of an EVCS.}
\label{fig:EVCS}
\end{figure}

In general, an EVCS acts as an interface or a high-wattage access point that capacitates the power exchange between a power grid and the EVs, thus enabling EV charging facilities. Fig.~\ref{fig:EVCS} presents the high-level overview of the CPS of an EVCS. The EVCS configured with AC bus, is provided electric power through the distribution grid topology system via step-down transformer. It possesses the capability to charge mutliple EVs simultaneously through multiple serving points on the bus. These connection points are the distributed EVCs which are conventional power outlets with specific current ratings. Also the EVCs are connected through the outgoing radial distribution feeders to the EVs.
A centralized battery energy storage system (BESS) is integrated into the LV DC link using an AC/DC converter which performs secondary services, such as energy arbitrage, and other grid ancillary services. 
Further, EVCS is equipped with the central EVCS management system (EVCSMS), which receives charging requests from EVCs and acts as a standard access point to discrete EVCs over Open Charge Point Protocol (OCPP). 
Moreover, OCPP authorizes open communication between a smart EVCS and the cloud-based backend where the system engineers can remotely upgrade firmware, bill customers, optimize charging, and other functions. Some EV communications with an EVCS are governed by message types defined by the IEC 61850 standard. 

\section{PROPOSED CYBERSECURITY FRAMEWORK}
\begin{figure}[t]
\centering
\includegraphics[width= .45\textwidth, height = 0.8 in]{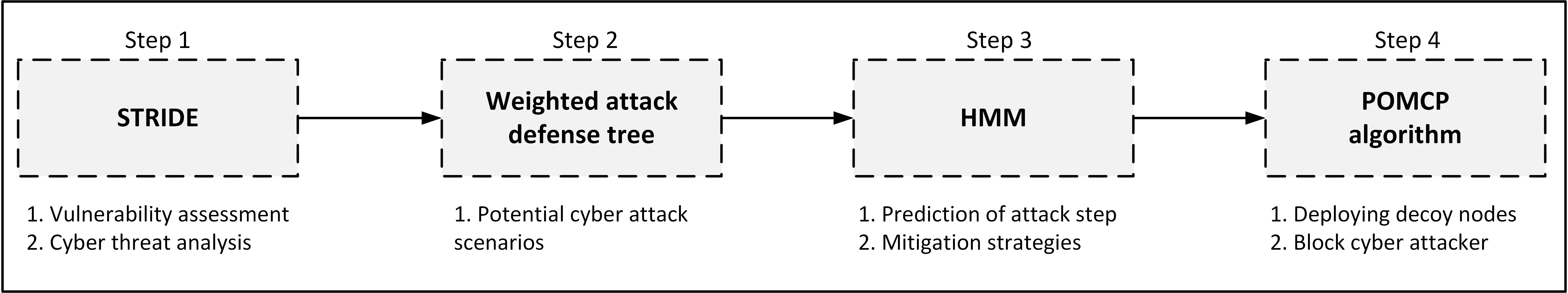}
\caption{Proposed cybersecurity framework.}
\label{fig:CS_framework}
\end{figure}

Fig.~\ref{fig:CS_framework} gives an overview of the cybersecurity framework of an EVCS. Step 1 evaluates the threat analysis using STRIDE, where existing vulnerabilities and cyber threats within an EVCS are identified. Step 2, i.e., weighted attack defense tree, studies diverse attack models which could be used by an attacker to target the system. 
Thus, the cyber attack scenarios are used as an input for HMM in step 3. HMM predicts the attack steps in a multi-step attack scenario by using viterbi algorithm and proposes various mitigation strategies. Finally, at step 4, a POMCP algorithm is employed, where decoy nodes are added along with the predicted path to deceive the attacker if (s)he attacks the system through an un-predicted path.

\subsection{STRIDE Threat Modeling}

A literature review has identified numerous threat modeling methodologies, e.g., system theoretic process analysis for security (STPA-sec), hazard and operability (HAZOP), security-aware hazard and risk assessment (SAHARA), process for attack simulation and threat analysis (PASTA), and STRIDE~\cite{12342351351}. STRIDE is centered around the identification and mitigation of potential cyber threats against each system entity~\cite{8260283}. 

It is a categorical threat analysis model and has been adopted in many previous works~\cite{8260283},~\cite{9202653}. This paper incorporated STRIDE to discover the potential susceptibilities or security risks associated with each component of an EVCS at the earlier stages of development. It consists of a data flow diagram (DFD), highlighting different assets, access points, information flow, and potential vulnerabilities in the charging environment. It specifies a mnemonic for six different types of security threats as described below: 


\begin{enumerate}
    \item \textbf{S}poofing: It indicates impersonating an authorized source or system entity by fabricating data. For e.g., gaining access to the EVC and installing unauthorized remote command sequences or malware. The malware injected in the EVC may propagate to the EVs or distribution system resulting in a temporary shutdown of EVCS~\cite{1231351351235}.
    \item \textbf{T}ampering: It causes the forging of legitimate information~\cite{151351235}. 
    For example, EVC might inject erroneous charging or discharging information into EVCSMS or various data exchanges (e.g., energy request, demand response (DR), DR request, EV ID, and utility ID) between EVs, EVSE, and EVCSMS could be distorted. Hence, successful tampering on an EVCS could damage EVs, EVSE, XMS, and even the distribution transformer by misleading the charging and grid operations.
    \item \textbf{R}epudiation: It refers to invalidating or suppressing the information exchange between the charging system components~\cite{151351235}. 
    For instance, an infected EVC might deny the charge processing fee from an EV owner, even it has been paid for the corresponding charging. Hence, it will cause a financial loss to the EV owner. 
    \item \textbf{I}nformation disclosure: It refers to the data infringement or data leakage causing privacy violations~\cite{151351235}. For instance, the wireless or wired communication between the EVC and EV is highly prone to be intercepted by the threat agent. Hence, the attacker might eavesdrop and gain an unauthorized access to the security-sensitive information. This might result in endangering the privacy of both EVSE and EV.  %
    \item \textbf{D}enial of Service: It refers to the inaccessibility of network services to legitimate users due to flooding of the bus network with malicious traffic injected by an attacker~\cite{151351235}. It might shut down the entire EVCS through the termination of the charging process of connected EVs. 
    
    
    \item \textbf{E}levation of privilege: It occurs when a malevolent threat agent acquires more authority or privileges to access data in a CPS as compared to the intended user~\cite{151351235}. 
    To exemplify,  
    if a malicious agent gets critical attributes of transmitted EVC data (e.g., firmware, EV ID, or user data), (s)he might corrupt this data and hamper the EVCSMS functions and modify system files or software configurations. As a result, an entire EVCS could be crippled due to data poisoning.
\end{enumerate}

\begin{figure}[t]
\centering
\includegraphics[width= 0.5\textwidth, height = 2.8 in]{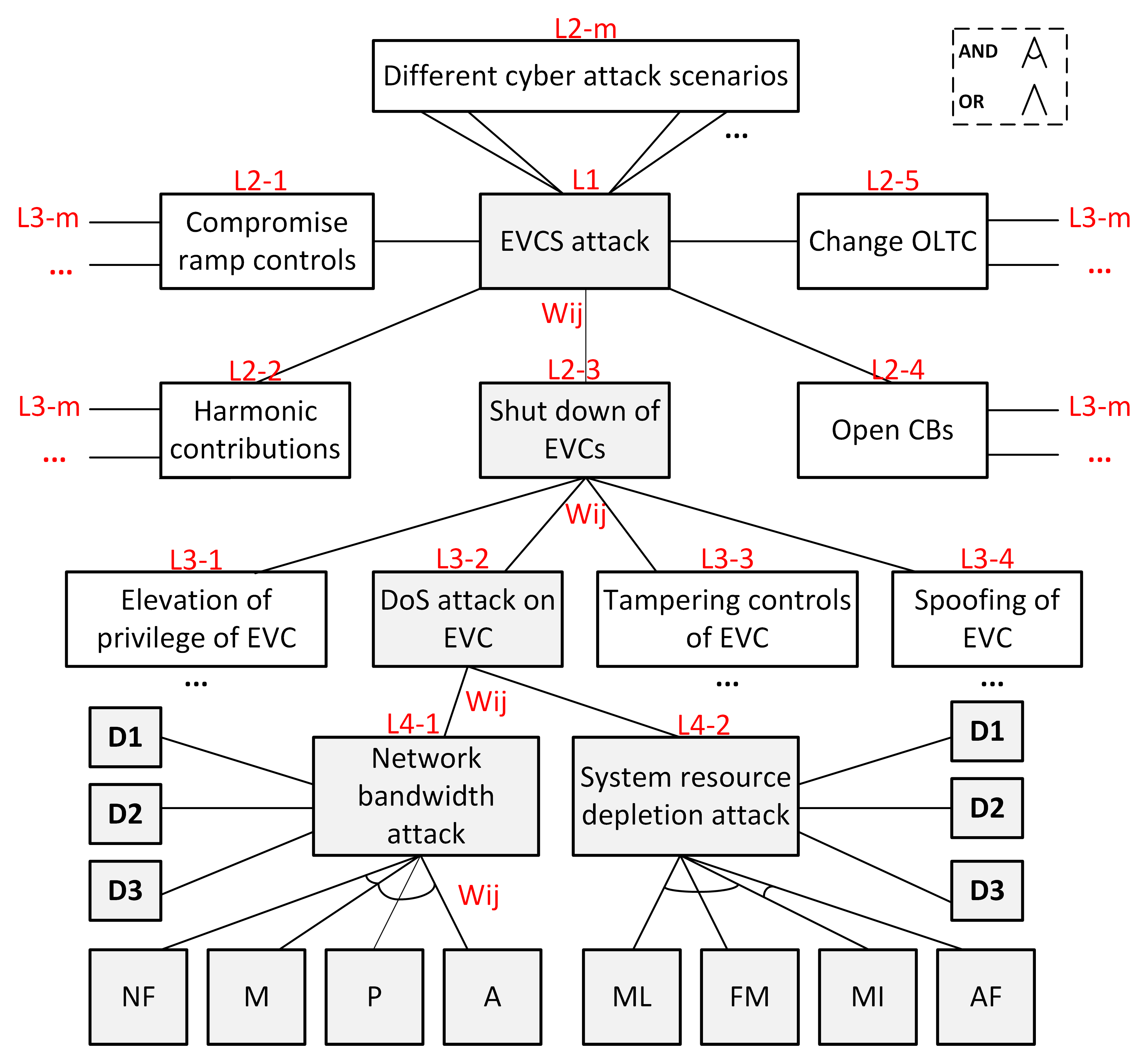}
\caption{An example of weighted attack tree for DoS.}
\label{fig:DoS_tree}
\end{figure}

\subsection{Weighted Attack Defense Tree}

The proposed weighted attack defense tree, as shown in Fig.~\ref{fig:DoS_tree}, analyses an adversary’s attack objectives by exhibiting a graphical tree notation in the form of a multi-level hierarchy process.
The root node (L1: EVCS attack) of an attack tree defines the primary goal of an attacker, which can be accomplished through different combinations of attack steps (or sub-goals, i.e., L2-2, L2-3, L2-4). 
Each attack leaf may constitute one or more defense nodes and logical “AND” and “OR” nodes. To exemplify, an attacker might accomplish the goal of group L2-3 to shut down EVCs by various cyber attack scenarios. One of the scenarios (i.e., L3-2) is DoS attack on the charger.
The agent might achieve its goal by attacking through two different attack scenarios or paths, e.g., network bandwidth attack (NB) (L4-1) and system resource depletion attack (SR) (L4-2). NB has three “AND” conditions, i.e., (1) network flooding (NF) and malformed packet (M), (2) M and protocol vulnerability exploitation (P), (3) P and amplification (A) attack. Also, SR has three “AND” conditions, i.e., (1) multiple login attempts (ML) and file system modification (FM), (2) FM and malware installation (MI), (3) MI and application flooding (AF). The attack tree also proposes several mitigation actions at each attack scenario (L4-1 and L4-2) for defense strategies. $W_{ij}$, a weighting factor, is used for defense cost (by an operator) or attack cost (by an attacker). 
It also considers other primary factors, e.g., defense cost $C_{ij}$ for developing defensive strategies in a CPS ~\cite{8870734}. A defender intends to establish minimum defense cost to secure the optimal attack nodes during the design process. Another parameter of paramount importance is the vulnerability index ($V_{k}$)~\cite{4275642}, which identifies the vulnerable assets in an EVCS. It varies from 0 (most vulnerable) to 1 (least vulnerable). 
Hence, the optimal defensive strategy (ODS) can be computed by minimizing $C_{ij}$ and maximizing $V_{k}$,

\begin{align}
\mathbf{ODS} = min \sum_{i=1}^{N} \sum_{j=1}^{M} C_{ij} + max \sum V_{k}.
\end{align}

\subsection{Hidden Markov Model}
HMM, a stochastic predictive model, is proposed as an intrusion detection and prevention system (ID\&PS). It models the interactions between the attacker and an EVCS for each multi-step attack scenario and predicts the subsequent attack steps to possibly prevent attacks and minimize the damage.

For the implementation of the proposed ID\&PS, multi-step attack files (attack steps of an attacker from the weighted defense attack tree) and historical cybersecurity data, as well as system logs, 
are used to train the proposed HMM by supervised machine learning. The learning module produces HMM configuration files which represent the state transition matrix, emission matrix, and the number of transition states, which are important aspects for the prediction module. Further, a real-time monitoring system produces a sequence of alerts and notifications indicating an intrusion. The module uses the HMM algorithm on the alert stream and assesses the attack steps, followed by taking appropriate mitigation actions. 

Mathematically, HMM is expressed as two probabilistic processes~\cite{8031986}: (1) a hidden stochastic process ($h(n)$), which corresponds to the state transition, and (2) an observation process ($o(n)$), which corresponds to the emitted observations from each hidden state. 


Hence, a discrete first-order HMM is given by
\begin{align}
\mathbf{\lambda} = (\Sigma, S, T, E, \pi).
\end{align}

Where ${\Sigma}$ represents a set of observations such that 
\begin{align}
\mathbf{\Sigma} = \{h_1, h_2, h_3, ..., h_M\}.
\end{align}


$\mathbf{S}$ represents a set of HMM states such that
\begin{align}
\mathbf{S} = \{S_1, S_2, S_3, ..., S_N\}.
\end{align}

$\mathbf{T}$ is an $N \times N$ state transition matrix for $N$ probable states, which describes probability of state transition ($t_{ij}$) from state $i$ to state $j$, such that


\begin{multline}
 \mathbf{T} = \{\{t_{ij}\}_{N \times N} \mid t_{ij} = P (q_{n+1} = S_j \mid q_{n} = S_i)   \},\\ \text{where } 1\leq i, j \leq N   
\end{multline}

$\mathbf{E}$ is an $N \times M$ emission matrix, which indicates the probability of different alerts ($e_j$) for a specific attack scenario such that 

\begin{multline}
\mathbf{E} = \{\{e_{j} (k)\} _{N \times M} \mid e_j = P (h_{k}\mid q_{n} = S_j) \}, \\ \text{where } 1 \leq j \leq N, 1\leq k \leq M.
\end{multline}

$\mathbf{\Pi}$ is an $N \times 1$ initial probability distribution vector, which indicates the probability of the initial state of the attack such that 
\begin{align}
\mathbf{\Pi} =\{\{\pi_i\} _{N \times 1} \mid \pi_i = P (q_1 = S_i) \}, \text{where } 1 \leq i \leq N.
\end{align}

In this paper, supervised training is used to obtain the above mentioned probability matrices ($\mathbf{T}$ and $\mathbf{E}$) and initial distribution vector ($\mathbf{\Pi}$). It is defined mathematically by the following equations,
\begin{align}
t_{ij} = \frac{t(i,j)}{ \sum_{\alpha=0}^{N} \alpha \cdot t(i,\alpha) },
\end{align}

\begin{align}
e_{j}(k) = \frac{f(j,b_k)}{\sum_{\alpha=0}^{M} f(j,b_k)}.
\end{align}

Post the computation of $\mathbf{T}$, $\mathbf{E}$ and $\mathbf{\Pi}$, viterbi algorithm is applied to the $\mathbf{T}$ to estimate the most likely the state sequence, $\mathbf{Q}* = \{q_1, q_2,..., q_k\}$, for the given alert sequence, $\mathbf{H} = \{h_1, h_2, ..., h_k\}$ based on the state probability for each attack step. The likelihood of the best state sequence can be evaluated by the following formula:
\begin{align}
arg \hspace{0.125cm} max\{P (H \mid B, \lambda)\} =  arg \hspace{0.125cm} max\{P (H,B \mid \lambda)\}.
\end{align}

Let $\mathbf{\delta}_t(i)$ be the maximal probability of state sequences at time n that ends in state $S_{i}$ and produces the  first n observations for the given model, denoted by the following formula:

\begin{align}
\delta_t(j) = max\{P(S_1, S_2, ..., S_n, h_1, h_2, ..., h_n, q_n = S_j\mid \lambda )\}.
\end{align}
Hence, viterbi algorithm calculates the best state sequence and specifies the best state $S_i$ at time $n$.


\subsection{Partially Observable Monte-Carlo Planning (POMCP) Algorithm}
A high-skilled attacker might not follow the predicted path and cause abnormal behaviors by attacking through alternative attack paths. Hence, to block these un-predicted paths, the POMCP algorithm is used to drift the attacker towards the predicted route~\cite{9387290}. Furthermore, decoy nodes can be deployed, so once the attacker breaks into the decoy system, the monitoring system will generate alerts, and the attacker will be directed towards the predicted path. Finally, the defender can undertake appropriate mitigation actions to block the attacker beforehand. In this case, the defenders can safeguard all attack paths based on the predicted probability index (P$^{\text{Pr}}_{\text{L}}$), where P is the probability estimate of an attack state, Pr is the priority given to the attack stage, and L is the ith layer (i = 1,2,3, ..). 

\section{CASE STUDY}
An  attack scenario is constructed where a DoS attack is used to compromise the EVCs. There is a sequence of eight attack states involved in the hidden layer, which constitutes: (1) M, (2) P, (3) NF, (4) (A, (5) ML, (6) MI, (7) FM, and (8) AF. In addition, there are two alert sequences in the second layer, (1) NB and (2) SR.

\begin{figure}[t]

\centering
\includegraphics[width= .35\textwidth, height = 5.0 in]{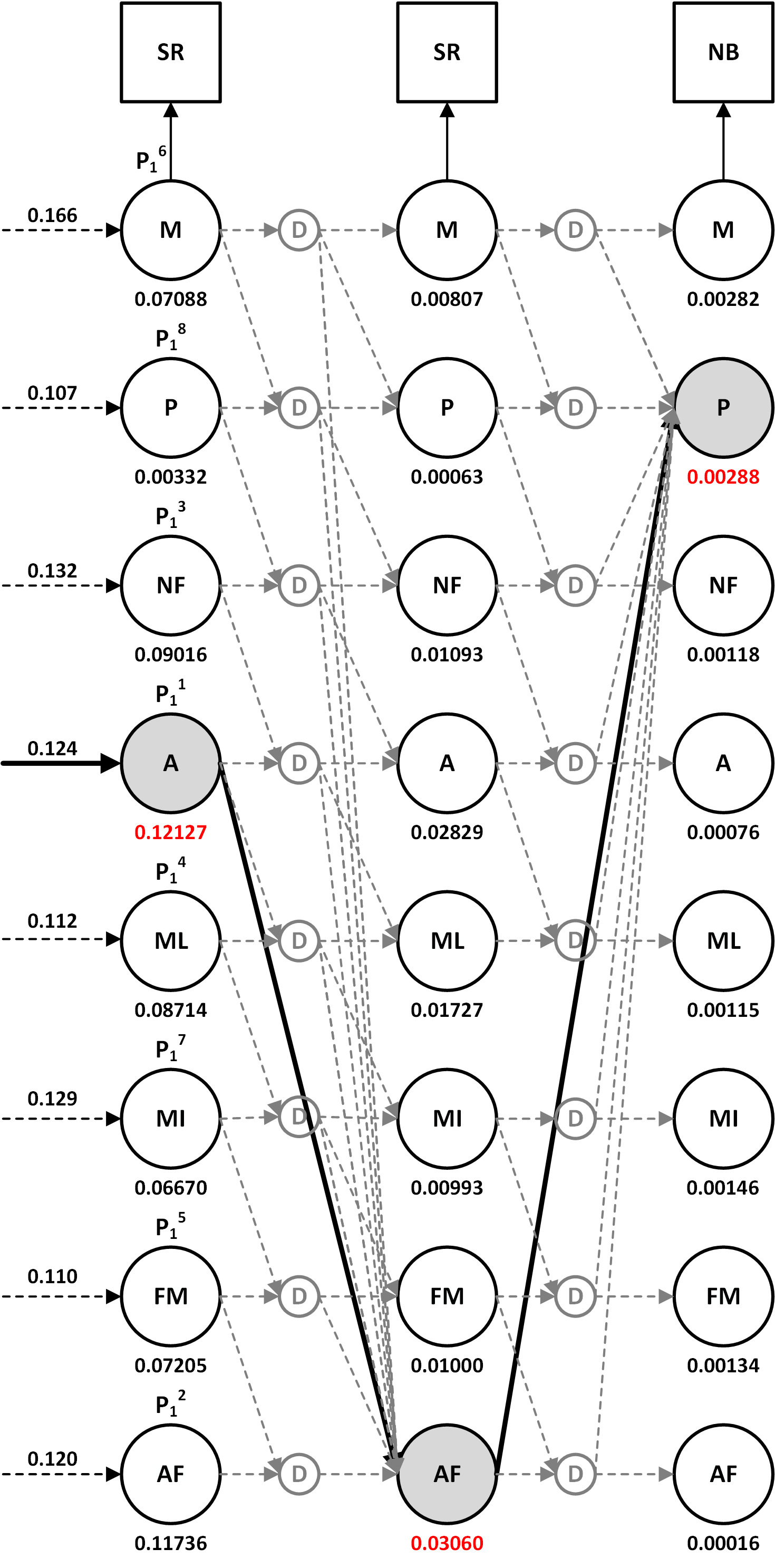}
\caption{A case study of Viterbi algorithm and POMCP algorithm.}
\label{fig:Case_study}
\end{figure}

In accordance with the viterbi algorithm, the system determines that the most likely attack path is “A”, “AF” and “P” when alert observations for “SR”, “SR” and “NB” are received from the monitoring system at three consecutive times. As shown by Fig.~\ref{fig:Case_study}, the most likely attack state (at $n$ = 1) is “A” with a maximum probability of 12.13\% for the SR alert, the most likely attack state (at $n$ = 2) is “AF” with a maximum probability of 3.06\% for the SR alert, and the most likely attack state (at $n$ = 3) is “P” with a maximum probability of 0.29\% for the NB alert. In other words, the threat agent tries to attack the EVCs by attempting SR attacks and NB attacks.


Further, as described earlier, an attacker may traverse through any attack path to reach its final goal. Assuming that there is an attack at the “A” attack stage, P$^{\text{1}}_{\text{1}}$=0.12127. Then the presence of decoy nodes (D) along with the predicted path nodes in the network deflects the attacker to the predicted path (i.e., A-AF-P). Now, even though the attacker is able to compromise the decoy node, the subsequent predicted nodes will deviate the attacker to another decoy in the system. Hence, the defenders could protect the network across each node.
 
Hence, the proposed cybersecurity model predicts cyber intrusions into an EVCS effectively as it renders explicit information on the current state of the attack. Further, the gained knowledge of attacks can be exploited to develop significant mitigation measures.

\section{POTENTIAL MITIGATION MEASURES}


This paper discusses various mitigation strategies to enhance the security and resiliency of an EVCS. 
For instance, different risk assessment and threat analysis models (e.g., STRIDE, OCTAVE) might be used in the early development stage to analyze multiple security threats and risks during the pre-attack stage.
Moreover, any unauthorized access to the malicious users can be impeded by implementing vulnerability scan, multi-authentication, signature-based cryptography of EVCS operation and communications.
Similarly, during the attack, the proposed HMM and POMCP can be engaged to study the current and future behavior of the cyber attacker.
However, for the post-attack mitigations, digital forensic analysis can be a useful tool to resume EVCS to its healthy state by identifying the compromised asset or vulnerabilities. 
Furthermore, system hardening can be implemented to reduce its attack surface by disconnecting the compromised components and using safer backup options. In addition, a system engineer can update the firmware or remove unauthorized software, set up an IDS\&IPS to reserve the issues.

\section{CONCLUSION}
This paper proposes a first-of-the-kind cybersecurity framework using STRIDE for designing vulnerability and risk assessment, a weighted attack defense tree to create attack scenarios, HMM, and POMCP algorithms to analyze the cyber threats and risks associated with an EVCS. Moreover, it also introduces potential mitigation strategies to enhance the security and resiliency of an EVCS. The proposed ID\&PS detects the abnormal behaviors, unauthorized intrusions, and anomalies of the threat actors. It focuses on predicting the attacker's potential targets and behaviors so that the defender can undertake active defense actions prior to the occurrence of cyber attacks.
Since the proposed framework is evaluated as an online decoding process, in the future, its real-time implementation can be evaluated and tested by designing a cyber-physical hardware-in-the-loop (HIL) testbed. Moreover, the effectiveness of the proposed mitigations can be studied under realistic and sophisticated cyber attack scenarios.

\bibliographystyle{IEEEtran}

\begin{thebibliography}{10}
\providecommand{\url}[1]{#1}
\csname url@samestyle\endcsname
\providecommand{\newblock}{\relax}
\providecommand{\bibinfo}[2]{#2}
\providecommand{\BIBentrySTDinterwordspacing}{\spaceskip=0pt\relax}
\providecommand{\BIBentryALTinterwordstretchfactor}{4}
\providecommand{\BIBentryALTinterwordspacing}{\spaceskip=\fontdimen2\font plus
\BIBentryALTinterwordstretchfactor\fontdimen3\font minus
  \fontdimen4\font\relax}
\providecommand{\BIBforeignlanguage}[2]{{%
\expandafter\ifx\csname l@#1\endcsname\relax
\typeout{** WARNING: IEEEtran.bst: No hyphenation pattern has been}%
\typeout{** loaded for the language `#1'. Using the pattern for}%
\typeout{** the default language instead.}%
\else
\language=\csname l@#1\endcsname
\fi
#2}}
\providecommand{\BIBdecl}{\relax}
\BIBdecl

\bibitem{4753673}
A.~N. {Bessani}, P.~{Sousa}, M.~{Correia}, N.~F. {Neves}, and P.~{Veríssimo},
  ``The crutial way of critical infrastructure protection,'' \emph{IEEE
  Security Privacy}, vol.~6, no.~6, pp. 44--51, 2008.

\bibitem{9361308}
J.~{Hong}, R.~{Karnati}, C.~W. {Ten}, S.~{Lee}, and S.~{Choi}, ``Implementation
  of secure sampled value (sesv) messages in substation automation system,''
  \emph{IEEE Transactions on Power Delivery}, pp. 1--1, 2021.

\bibitem{6786500}
J.~{Hong}, C.~{Liu}, and M.~{Govindarasu}, ``Integrated anomaly detection for
  cyber security of the substations,'' \emph{IEEE Transactions on Smart Grid},
  vol.~5, no.~4, pp. 1643--1653, 2014.

\bibitem{9306781}
R.~{Zhu}, C.~C. {Liu}, J.~{Hong}, and J.~{Wang}, ``Intrusion detection against
  mms-based measurement attacks at digital substations,'' \emph{IEEE Access},
  vol.~9, pp. 1240--1249, 2021.

\bibitem{7553505}
Y.~{Yang}, H.~{Xu}, L.~{Gao}, Y.~{Yuan}, K.~{McLaughlin}, and S.~{Sezer},
  ``Multidimensional intrusion detection system for iec 61850-based scada
  networks,'' \emph{IEEE Transactions on Power Delivery}, vol.~32, no.~2, pp.
  1068--1078, 2017.

\bibitem{9272723}
S.~{Acharya}, Y.~{Dvorkin}, H.~{Pandžić}, and R.~{Karri}, ``Cybersecurity of
  smart electric vehicle charging: A power grid perspective,'' \emph{IEEE
  Access}, vol.~8, pp. 214\,434--214\,453, 2020.

\bibitem{7807218}
D.~T. {Hoang}, P.~{Wang}, D.~{Niyato}, and E.~{Hossain}, ``Charging and
  discharging of plug-in electric vehicles (pevs) in vehicle-to-grid (v2g)
  systems: A cyber insurance-based model,'' \emph{IEEE Access}, vol.~5, pp.
  732--754, 2017.

\bibitem{12342351351}
W.~Young and N.~G. Leveson, ``An integrated approach to safety and security
  based on systems theory,'' \emph{Communications of the ACM}, vol.~57, no.~2,
  Feb. 2014.

\bibitem{8260283}
R.~{Khan}, K.~{McLaughlin}, D.~{Laverty}, and S.~{Sezer}, ``Stride-based threat
  modeling for cyber-physical systems,'' in \emph{2017 IEEE PES Innovative
  Smart Grid Technologies Conference Europe (ISGT-Europe)}, 2017, pp. 1--6.

\bibitem{9202653}
P.~{Danielis}, M.~{Beckmann}, and J.~{Skodzik}, ``An iso-compliant test
  procedure for technical risk analyses of iot systems based on stride,'' in
  \emph{2020 IEEE 44th Annual Computers, Software, and Applications Conference
  (COMPSAC)}, 2020, pp. 499--504.

\bibitem{1231351351235}
S.~Lightman and T.~Brewer, ``Symposium on federally funded research on
  cybersecurity of electric vehicle supply equipment (evse),'' \emph{Nat. Inst.
  Standards Technol. NISTIR 8294}, April 2020.

\bibitem{151351235}
D.~C. K.~Harnett, B.~Harris and G.~Watson, ``{DOE/DHS/DOT} volpe technical
  meeting on electric vehicle and charging station cybersecurity,'' \emph{U.S.
  Department of Transportation, Tech. Report. DOT-VNTSC-DOE-18-01}, Mar. 2018.

\bibitem{8870734}
B.~{Xu}, Z.~{Zhong}, and G.{He}, ``A minimum defense cost calculation method
  for attack defense trees,'' \emph{Security and Communication Networks}, vol.
  2020, pp. 1--12, 2020.

\bibitem{4275642}
C.-W. Ten, C.-C. Liu, and M.~Govindarasu, ``Vulnerability assessment of
  cybersecurity for scada systems using attack trees,'' in \emph{2007 IEEE
  Power Engineering Society General Meeting}, 2007, pp. 1--8.

\bibitem{8031986}
P.~{Holgado}, V.~A. {Villagrá}, and L.~{Vázquez}, ``Real-time multistep
  attack prediction based on hidden markov models,'' \emph{IEEE Transactions on
  Dependable and Secure Computing}, vol.~17, no.~1, pp. 134--147, 2020.

\bibitem{9387290}
M.~A. R.~A. Amin, S.~Shetty, L.~Njilla, D.~K. Tosh, and C.~Kamhoua, ``Hidden
  markov model and cyber deception for the prevention of adversarial lateral
  movement,'' \emph{IEEE Access}, vol.~9, pp. 49\,662--49\,682, 2021.

\end{thebibliography}

\end{document}